\documentclass{article}

\usepackage{arxiv}

\usepackage[utf8]{inputenc} 
\usepackage[T1]{fontenc}    
\usepackage{hyperref}       
\usepackage{url}            
\usepackage{booktabs}       
\usepackage{amsfonts}       
\usepackage{nicefrac}       
\usepackage{microtype}      
\usepackage{lipsum}
\usepackage{graphicx}
\graphicspath{ {./images/} }

\title{Inter-organisational patent opposition network: How companies form adversarial relationships}

\author{
 Tomomi Kito \\
  Faculty of Science and Engineering\\
  Waseda University\\
  3-4-1 Okubo, Shinuku-ku, Tokyo 169-8555 Japan\\
  \texttt{kito@waseda.jp} \\
   \And
 Nagi Moriya\\
    Waseda Innivation Laboratory\\
    Waseda University\\
    1-6-1 Nishiwaseda, Shinjuku-ku, Tokyo 169-8050 Japan \\
  \And
Junichi Yamanoi\\
   Waseda Innivation Laboratory\\
    Waseda University\\
    1-6-1 Nishiwaseda, Shinjuku-ku, Tokyo 169-8050 Japan \\
}

\begin{document}
\maketitle
\begin{abstract}
Much of the research on networks using patent data focuses on citations and the collaboration networks of inventors, hence regarding patents as a positive sign of invention. However, patenting is, most importantly, a strategic action used by companies to compete with each other. This study sheds light on inter-organisational adversarial relationships in patenting for the first time. We constructed and analysed the network of companies connected via patent opposition relationships that occurred between 1980 and 2018. A majority of the companies are directly or indirectly connected to each other and hence form the largest connected component. We found that in the network, many companies disapprove patents in various industrial sectors as well as those owned by foreign companies. The network exhibits heavy-tailed, power-law-like degree distribution and assortative mixing, making it an unusual type of topology. We further investigated the dynamics of the formation of this network by conducting a temporal network motif analysis, with patent co-ownership among the companies considered. By regarding opposition as a negative relationship and patent co-ownership as a positive relationship, we analysed where collaboration may occur in the opposition network and how such positive relationships would interact with negative relationships. The results identified the structurally imbalanced triadic motifs and the temporal patterns of the occurrence of triads formed by a mixture of positive and negative relationships. Our findings suggest that the mechanisms of the emergence of the inter-organisational adversarial relationships may differ from those of other types of negative relationships hence necessitating further research.
\end{abstract}


\section{Introduction}
\label{sec:1}
Innovation is one of the key issues in the recent economic paradigm and has consequently attracted a growing number of academic studies over time.
Enhanced by recent computational advances, patent statistics derived from large-scale patent databases have been extensively used to help gain insight into innovation processes, drivers, and appropriate measures for innovation values~\cite{Kogan 2017,Youn 2015,Hasan 2010}.

Network science has significantly contributed to patent data science. Considerable effort has been devoted to analysing networks of patents connected by citations, with the intent to understand technological change and impact~\cite{Erdi 2013,Verspagen 2007}. The consensus here is that the patent citation network structure reflects the evolutionary pattern of human technology. The measures for the technological importance of a patented invention are mostly based on how frequently the patent is cited by subsequent patents (e.g., \cite{Funk 2017,Gao 2018,Acemoglu 2016}).
Another stream of patent network research is on the social structure of collaboration among patentees. Through the analyses of co-applicant patent networks, various factors that may contribute to the production of high-quality inventions have been identified. These factors include the proximity of geographical locations and technological fields of collaborating organisations, their positioning in the inventors' networks, and so on~\cite{Balconi 2004,Fleming 2007,Graf 2011}. 

These studies consider patents as a positive sign of invention, which is a generally acknowledged view. However,patenting is, most fundamentally, a strategic action used by applicants to gain legal right in order to exclude others from making, using or selling an invention. While inter-organisational relationships can be intermittently friendly (e.g., R\&D collaboration and patent co-ownership), they are mostly adversarial/rivalry. This (rather essential) aspect of patenting has not received much attention in network science, and data science in general, to date. In management science, the strategic management of patents has been increasingly considered as the core for enhancing the competitiveness of companies. Considerable research has been conducted on the development of theoretical frameworks, case studies on the global patent wars, and analysis of the financial impact of strategic actions (for summary see Refs.~\cite{Somaya 2012,Grzegorczyk 2020}). However, the existing research does not go beyond treating the dyadic relationships between companies. It further provides no view on the network of companies connected by strategic actions. 

This study illuminates patent opposition as a company's key strategic action against others. Patent opposition is a legal action that a company (or individual) can take to challenge the validity of a patent within a certain period (usually 6--9 months) after grant. If an opposition is `successful', the opposed patent is revoked and cannot take effect in any of the signatories. Therefore, companies thus oppose patents owned by rival companies clearly intending to hinder their innovation activities~\cite{WIPO 2018,Sterlacchini 2016,Harhoff 2003}. 
In this study, we construct the patent opposition network where the nodes represents companies, rather than patents, and the edges represent oppositions. We analyse the properties of this network, with the aim of gaining insights into how companies may form adversarial relationships between each other. 
In social network analysis, although the main stream of research addresses networks with positive relationships (e.g., friendship and co-authorship), the importance of negative relationships (e.g., interpersonal dislike, conflict, and social exclusion) has attracted more interest~\cite{Duffy 2002,Labianca 2006,Venkataramani 2013}. Scholars have pointed out distinct features of negative relationship networks that are fundamentally different from positive relationship networks. Such features include the low connectivity among nodes, and a low level of transitivity -- that is, while friends of friends are usually friends in positive relationship networks, the enemies of enemies are not necessarily enemies in negative relationship networks~\cite{Everett 2014}. However, it is yet to be studied whether such insights derived from the analyses of social relationships are applicable to inter-organisational adversarial networks. We examine this by analysing the patent opposition network. 

Furthermore, we consider patent co-ownership among companies that have been involved in opposition. We investigate how the companies' strategies for ``suppressing the inventions of other companies''---captured by oppositions---and ``accelerating inventions''---captured by  co-ownership---may interact. In social network analysis, theories have been developed to discuss how positive and negative relationships might evolve together. The main theory is the structural balance theory~\cite{Heider 1946}~\cite{Cartwright 1956}, which states that when a triad is a signed graph (i.e., a graph in which each edge has a positive or negative sign), the structure is balanced if the multiplicative product of the signs of the edges is positive and imbalanced if this product is negative. 
In other words, the stability of the various triads conforms to the following simplified social principles: (1) My friend's friend is my friend, (2) my friend's enemy is my enemy, (3) my enemy's friend is my enemy, and (4) my enemy's enemy is my friend. Here, positive and negative relationships are interpreted as being friends and enemies, respectively. The balance theory suggests that balanced rather than imbalanced triads will become more frequent over time. In other words, the number of instances of the triadic relational patterns (1)--(4) would increase over time. However, how such balanced structure may be achieved remains insufficiently studied. Although some scholars have investigated the dynamic interplay between positive and negative relationships in online social media networks~\cite{Leskovec 2010}, there are enough reasons to assume that the properties of strategically formed relationships among companies would be substantially different from the properties of human social interactions~\cite{Labianca 2014}. We apply temporal dyadic and triadic motif analysis to our data to gain insights into the dynamic formation of adversarial and collaborative relationships in the inter-organisational network. 

The rest of this paper is organised as follows. Section~\ref{sec:2} gives a detailed description of the rationale of focusing on patent opposition and the legal background of opposition in more detail. It also explains the procedures through which we collected and reformatted the data for our analysis. 
Sections~\ref{sec:3} and \ref{sec:4} describe our main analysis of the inter-organisational patent opposition network. Section~\ref{sec:3} investigates the global pattern of the network, while Section~\ref{sec:4} investigates how such a global pattern may have been formed by conducting the temporal motif analysis. Section~\ref{sec:5} summarises our findings and concludes the paper.

\section{Patent Opposition and Data}
\label{sec:2}
\subsection{Patent opposition and databases}
\label{subsec:2.1}
Patent opposition is a legal means that allows third parties to challenge the validity of a patent within a given period , usually 6--9 months, after a grant. However, this period varies with jurisdiction. When an opposition is made, the validity of the opposed patent is decided by the patent office rather than courts. There are three possible outcomes: the patent is maintained in its current form, amended, or revoked. Besides revocation, amendment can also be considered as a victory for the opposing party, since it impels the narrowing of the scope of the challenged patent~\cite{WIPO 2018}. Making an opposition is clearly an adversarial action of a company against its competitors as it is considerably costly.

While litigation is also a legal means for a company to render the patents of other companies invalid, it is substantially different from opposition. Litigation is a legal dispute that cannot not be settled by the parties involved out of court and therefore requires adjudication~\cite{Somaya 2012}. The costs incurred during litigation are several orders of magnitude higher than those during opposition. For example, in the case of the European Patent Office (EPO), oppositions relatively cost around US\$45,000~\cite{Chien 2018}, whereas litigation can cost above US\$ 1 million~\cite{AIPLA 2017}. Consequently, compared to opposition, litigation is an uncommon event~\cite{Somaya 2012}, and a possible option for large companies only. Other types of business entities that may initiate litigation are those referred to as ``trolls''. They do not engage in production or research and development (R\&D) themselves, but rather acquire patents from failed companies (or independent innovators) and assert them against producing entities to win court judgments for profit~\cite{WIPO 2018}. The proliferation of trolls has recently become a major concern, as they cause non-negligible impact on invention~\cite{Cohen 2017}. However, even with legal expertise, it is very difficult to detect and block their activities.

Since this study primarily aims to investigate how companies are connected via adversarial and collaborative relationships in patenting, we limit our focus only on patent opposition and joint ownership, but not litigation. By doing so, we can include small- and mid-sized companies that may play key roles, and exclude the influence of trolls that are somewhat insulated from strategic interrelations among inventing companies. 

While legal experts and practitioners have long discussed the importance and effectiveness of the opposition processes~\cite{Merges 1990,Farrell 2004}, economic literature has focussed more on the determinants of opposition. 
As the most significant determinant, empirical evidence converges in pointing to patent value. Simply put, the most valuable patents are more likely to be opposed. Companies known for being innovative tend to own opposed patents far more often than on average~\cite{Harhoff 2003}. It has been ascertained that the monetary value of the patent is positively correlated with the forward citation count, which is the number of subsequent patents that cite the patent~\cite{Harhoff 1999,Lanjouw 2001}. The forward citation is thus considered as the most effective indicator for patent value. Further, some scholars have also asserted that it significantly contributes to increasing the opposition probability~\cite{Harhoff 2004,van Zeebroeck et al 2011}. 
These studies provide some insight into the characteristics of opposed patents and their owners, rather than on how companies may be connected via the oppositions.

In this study, we collected the data on opposed patents and owner companies, and constructed a network of companies interrelated through opposition. For this purpose, we obtained data from the ``Orbis Intellectual Property Database (Orbis IP)''~\cite{Orbis IP} and ``Orbis Database (Orbis)''~\cite{Orbis}, both provided by Bureau van Dijk who is one of the most major publishers of business information. Orbis IP contains information on approximately 115 million patents worldwide, such as publication information, ownership, industry, history of transfer, and opposition. Orbis is a database that contains more than 360 million (mostly private) companies.
Companies that appear in the Orbis IP database are linked to those registered in the Orbis database via the same ID. Simultaneously using these two databases allowed us to identify various companies and their opposition and collaboration relationships.

\subsection{Data preparation}
\label{subsec:2.2}
\begin{figure}
    \begin{center}
          \includegraphics[width=11cm]{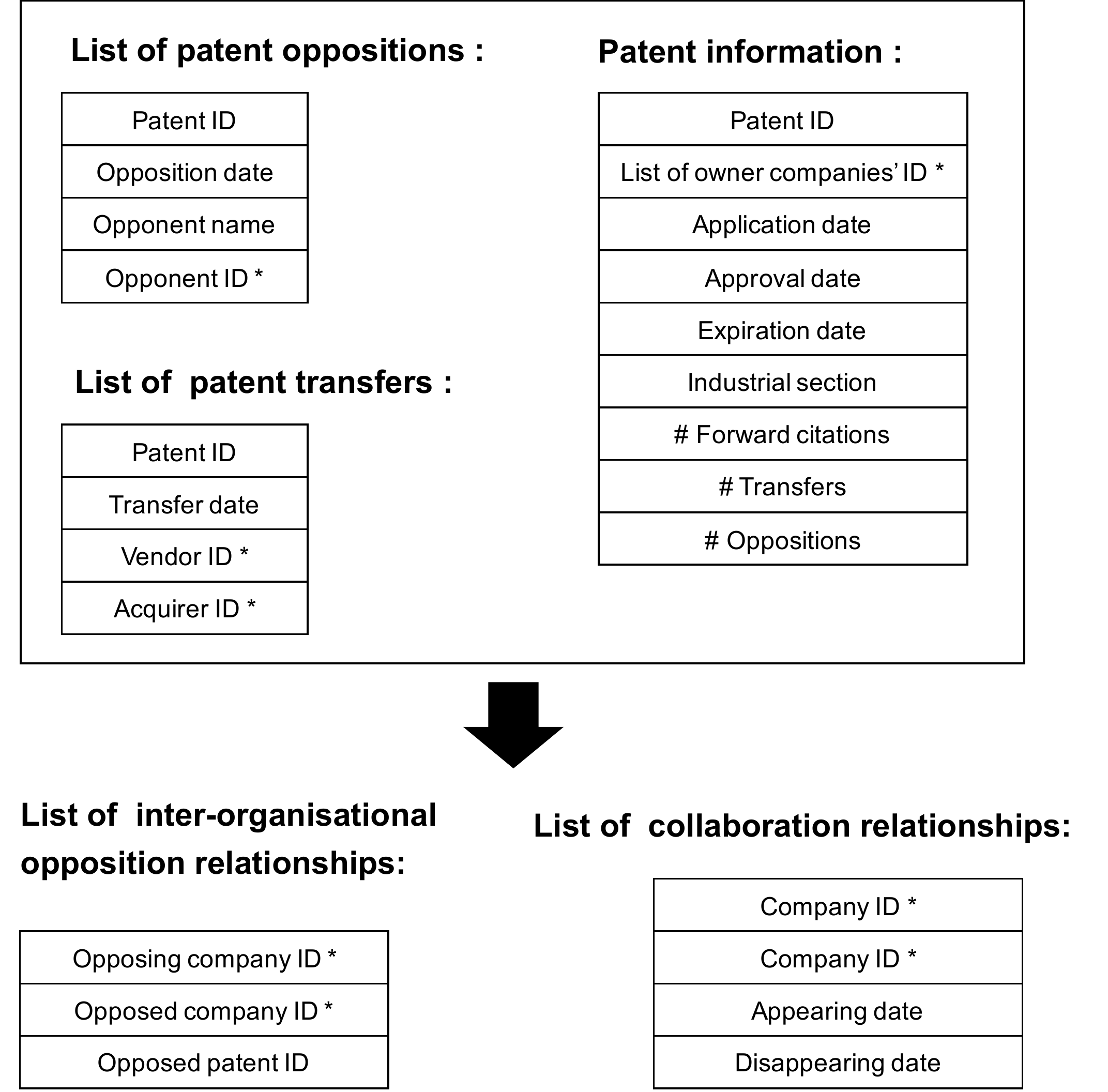}
        \caption{Data structure and data items used in the study}
        \label{fig: data structure}   
    \end{center}
\end{figure}

\begin{figure}
 \begin{center}
      \includegraphics[width=10cm]{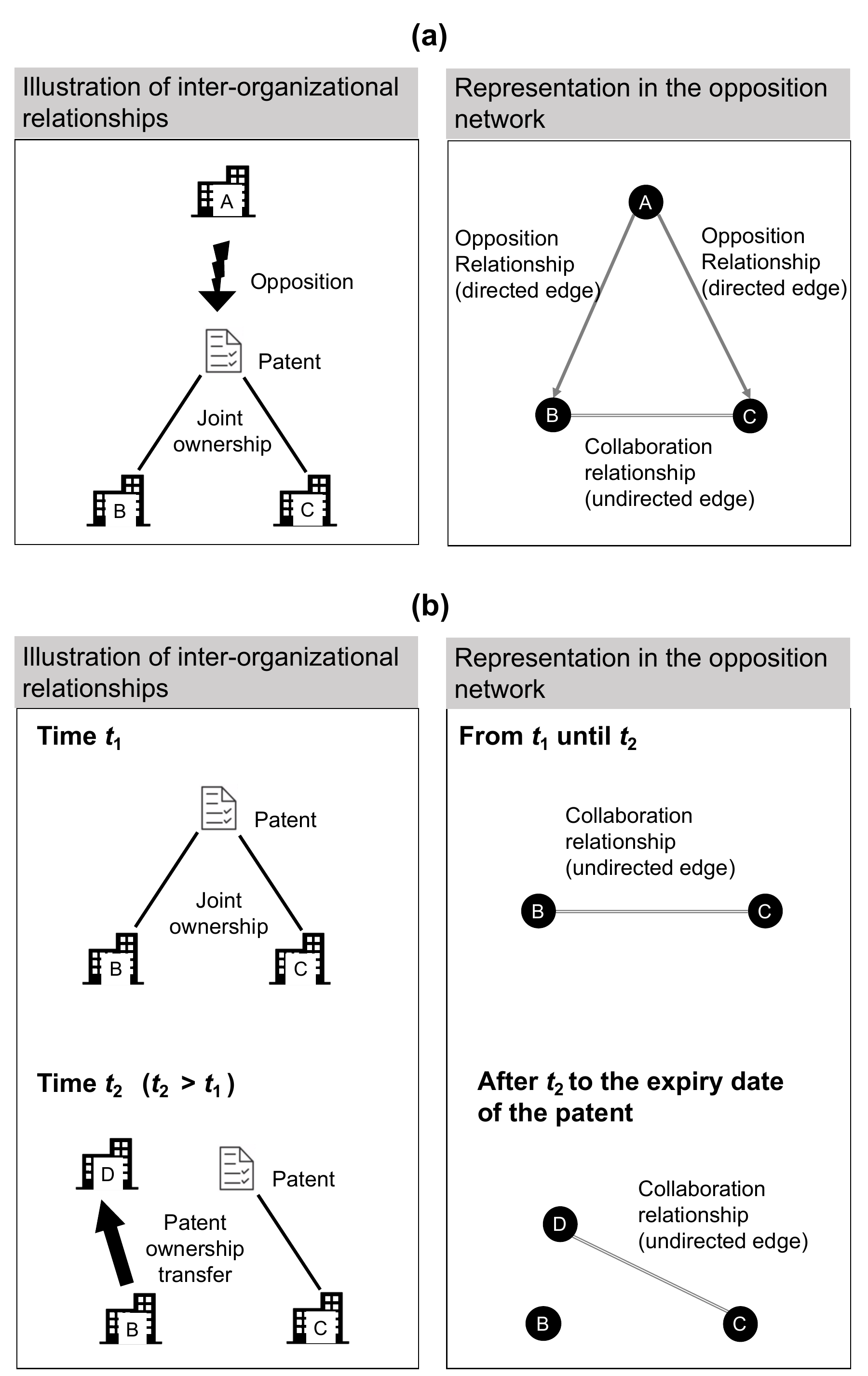}
      \caption{Rules of edge formation for the construction of the opposition network
       \protect \newline {\it Notes: (a) The left panel explains the situation in which Company A opposes a patent jointly owned by companies B and C. This situation is represented in the opposition network as directed opposition edges formed from A to B and from A to C, and a collaboration edge formed between B and C (illustrated in the right panel). (b) The left panel describes the situation in which companies B and C start owning a joint patent at $t_1$, and at some time later $t_2$ B's patent right was transferred to another company D. This situation is represented in the network as a collaboration edge between B and C from $t_1$ to $t_2$, and another collaboration edge between D and C after $t_2$ until the patent's expiry date.}}
      \label{fig: edge formation}  
       \end{center}
\end{figure}

This section describes the method we used to acquire and pre-process data for the analysis of the inter-organisational opposition network.
We extracted data from Orbis and Orbis IP and reformatted it as follows. Figure~\ref{fig: data structure} describes the data structure. In the opposition network, nodes and directed edges represent companies and opposition relationships respectively. Collaboration relationships among companies are also captured as undirected edges. Figure~\ref{fig: edge formation} illustrates the rules through which we created these two types of edges.

\begin{enumerate}
  \item We extracted the list of all the oppositions made between 1980 and 2018 from the Orbis IP database. Each opposition case was tagged to a patent ID,  opposition date, name and ID of the opponent company  (i.e. the company that made the opposition). The company ID is blank if the company is not registered in the Orbis company database.
  \item We then extracted all the patents that appeared in the list of oppositions and identified 31,470 patents. We only used the items listed in Fig.~\ref{fig: data structure}. We identified all the co-patenting companies if the patent is jointly owned by several companies. The numbers of forward citations, transfers and oppositions were counted as of the end of 2018.
  \item We also extracted the list of events of patent transfer for the patents that appeared in the list of patent oppositions. The list contains information on the patent ownership that was transferred from which company (i.e., vendor) to which company (i.e., acquirer), and the date on which the transfer occurred (i.e., the transfer date). 
  \item We searched all the company IDs (marked with $\ast$ in Fig.~\ref{fig: data structure}) in the Orbis company database and consequently identified company pairs (opposing and opposed companies) for each patent opposition. In the case where an opposed patent was jointly owned by several companies, we created a relationship between the opponent and each of the owner companies.
 In the analysis, the companies whose IDs were not found in the Orbis database were neglected. Through this procedure, we identified 11,480 companies and 26,433 opposition relationships among them. 
 \item We also obtained collaboration relationships among companies that were involved in opposition. For simplicity, we defined collaboration as patent co-ownership. A collaborative relationship between a pair of companies appears when the patent co-owned by the two companies was approved (i.e., the appearing date), and disappeared when expired or transferred to another company (or companies). Patent ownership transfer can occur when the patent right is acquired by another company or when the owner company sells the patent right. In such cases, the original collaboration relationship disappears, and a new collaboration relationship is formed with the new owner (or, each of the new owners in a case with more than one new owner). We identified 1,554 unique collaboration relationships.
 \item Based on the list of inter-organisational opposition relationships (see Fig. \ref{fig: data structure}), we created the opposition network, a network of companies connected via opposition relationships. In the network, nodes and directed edges represent companies and opposition relationships respectively.
In other words, if company A opposed a patent owned by company B, an opposition edge was created from company A pointing to company B.
Additionally, we added undirected edges representing collaboration relationships.
\end{enumerate}

Regarding each patent's industrial section, the information on the IPC (International Patent Classification) code was available. The IPC represents the whole body of technical knowledge that may be considered suitable to the field of patents for innovation, and is divided into the following eight sections: 
(A) HUMAN NECESSITIES, (B) PERFORMING OPERATIONS, TRANSPORTING, (C) CHEMISTRY, METALLURGY, (D) TEXTILES, PAPER, (E) FIXED CONSTRUCTIONS, (F) MECHANICAL ENGINEERING, LIGHTING, HEATING, WEAPONS, BLASTING, (G) PHYSICS, (H) ELECTRICITY. The sections are the highest level of hierarchy of the classification. We used this section information in this study. Each patent is tagged to one of these sections.

\subsection{Characteristics of opposed patents}
\label{subsec:2.3}
\begin{figure}
\begin{center}
      \includegraphics[width=10cm]{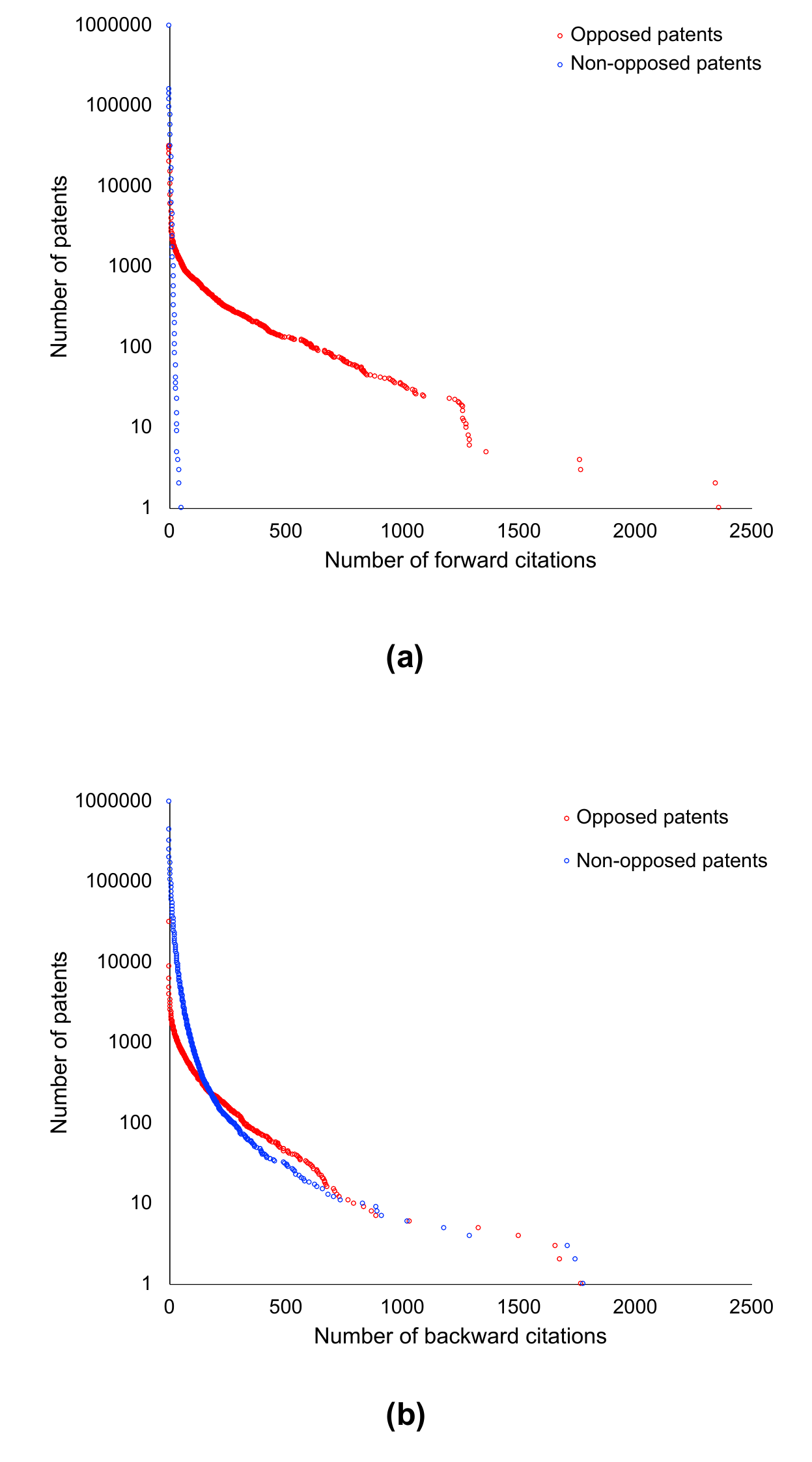}
      \caption{Cumulative distribution of the numbers of (a) forward citations and (b) backward citations for opposed and non-opposed patents}
      \label{fig: citations}   
      \end{center}
\end{figure}

Prior to the analysis, we checked the characteristics of the opposed patents in our data. Figure~\ref{fig: citations} shows the cumulative distributions of the numbers of forward (Figure (a))and backward citations (Figure (b)) for opposed and non-opposed patents. Here, non-opposed patents totalled 1 million patents with no record of oppositions, sampled uniformly at random from the Orbis IP database. The y-axis is shown on a logarithmic scale. Figure (a) indicates that the distributions of the number of forward citations are clearly different between the opposed and non-opposed patents. The maximum number of forward citations of non-opposed and opposed patents was 56 and 2,362 respectively. Although not all the opposed patents attract citations, the claim made by the literature stating that the number of forward citations indicates the higher likelihood of opposition seems to hold true~\cite{Harhoff 2004,van Zeebroeck et al 2011}. 
As explained in Section~\ref{subsec:2.1}, a high number of forward citations is regarded as a sign of a high patent value. Oppositions have been made against patents with many forward citations (high values). This fact may differentiate the adversarial relationship based on opposition from that based on mere negative feelings (e.g., dislike) in social networks.

In contrast, Figure~\ref{fig: citations} (b) shows that there is no clear difference in the distribution of the number of backward citations between opposed and non-opposed patents. Some scholars reported that they did not find any impact of backward citations on the likelihood of an opposition~\cite{Sterlacchini 2016} through their analysis of patent data from the EPO. We found that this holds true even when patents registered at other patent offices are taken into account.

\section{Inter-organisational Opposition Network}
\label{sec:3}
In this section, we analyse the inter-organisational patent opposition network, created via the sixth step of the procedures described in Section~\ref{subsec:2.2}. 
In this network, the nodes and edges represent companies and opposition relationships that existed at some point in time, respectively. The network comprises of 11,480 nodes (companies) and 26,433 directed edges (opposition relationships). The network is a directed multigraph with no loops. In other words, two nodes may be connected by more than one directed edge in the same direction. This happens when a company opposes multiple patents that are owned by the same company. The number of unique pairs of nodes that were connected by one or more edge(s) (i.e., the number of edges when disallowing multiple edges) was 14,320.

\subsection{Largest connected component}
\label{subsec:3.1}
\begin{figure}
\begin{center}
      \includegraphics[width=11.8cm]{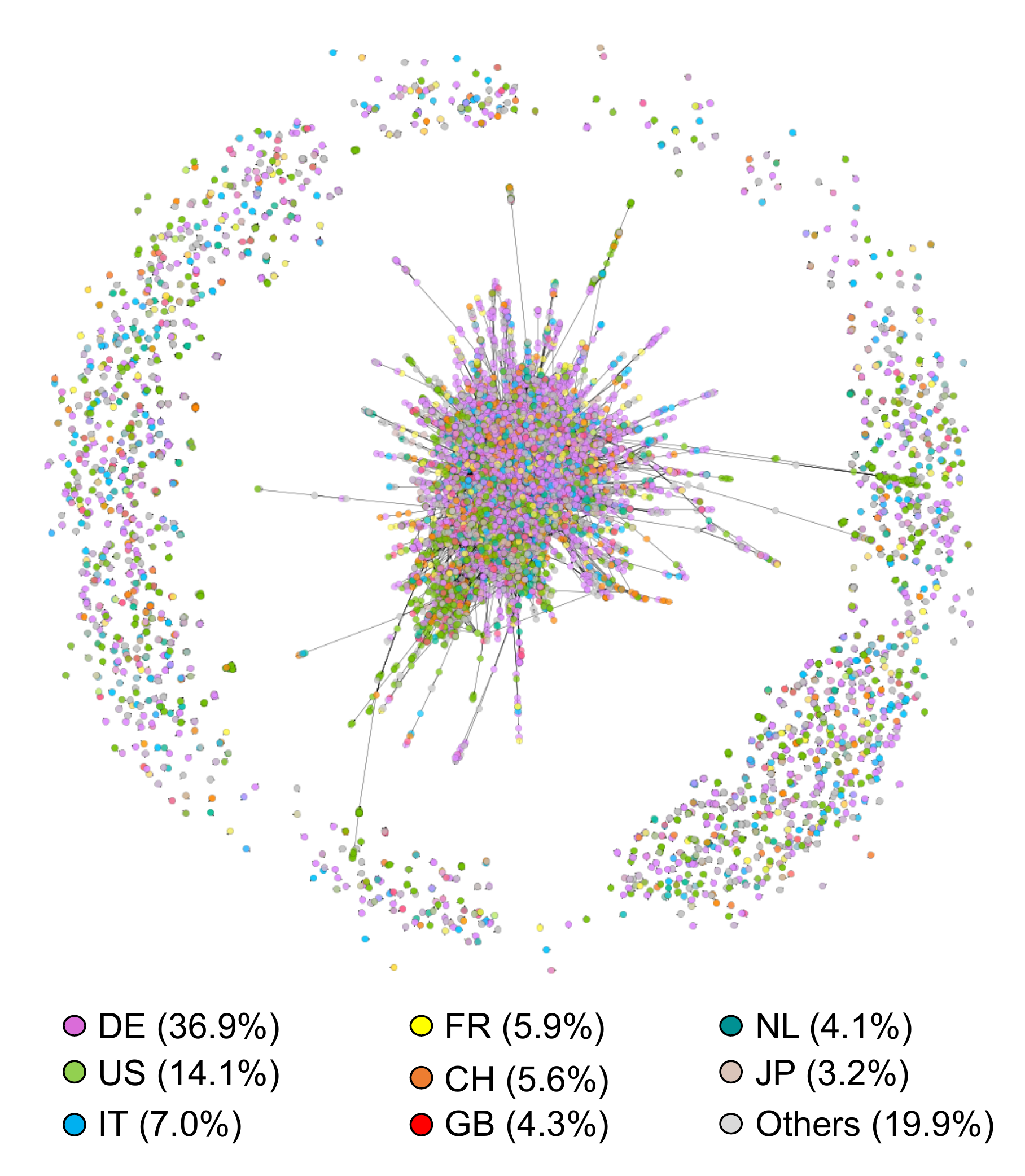}
      \caption{Inter-organisational opposition network.
      \protect \newline {\it Notes: Nodes represent companies, and edges represent opposition relationships. Regarding the visualisation, a force-directed layout algorithm~\cite{ForceAtlas2} was applied, making nodes that are tightly connected move close to each other. Node colour represents the country in which the company is located, identified by the country code provided by the Orbis database. The percentage (shown in brackets) is the ratio of the number of the companies in the region.}}
      \label{fig: opposition network}   
      \end{center}
\end{figure}

\begin{figure}
\begin{center}
      \includegraphics[width=10cm]{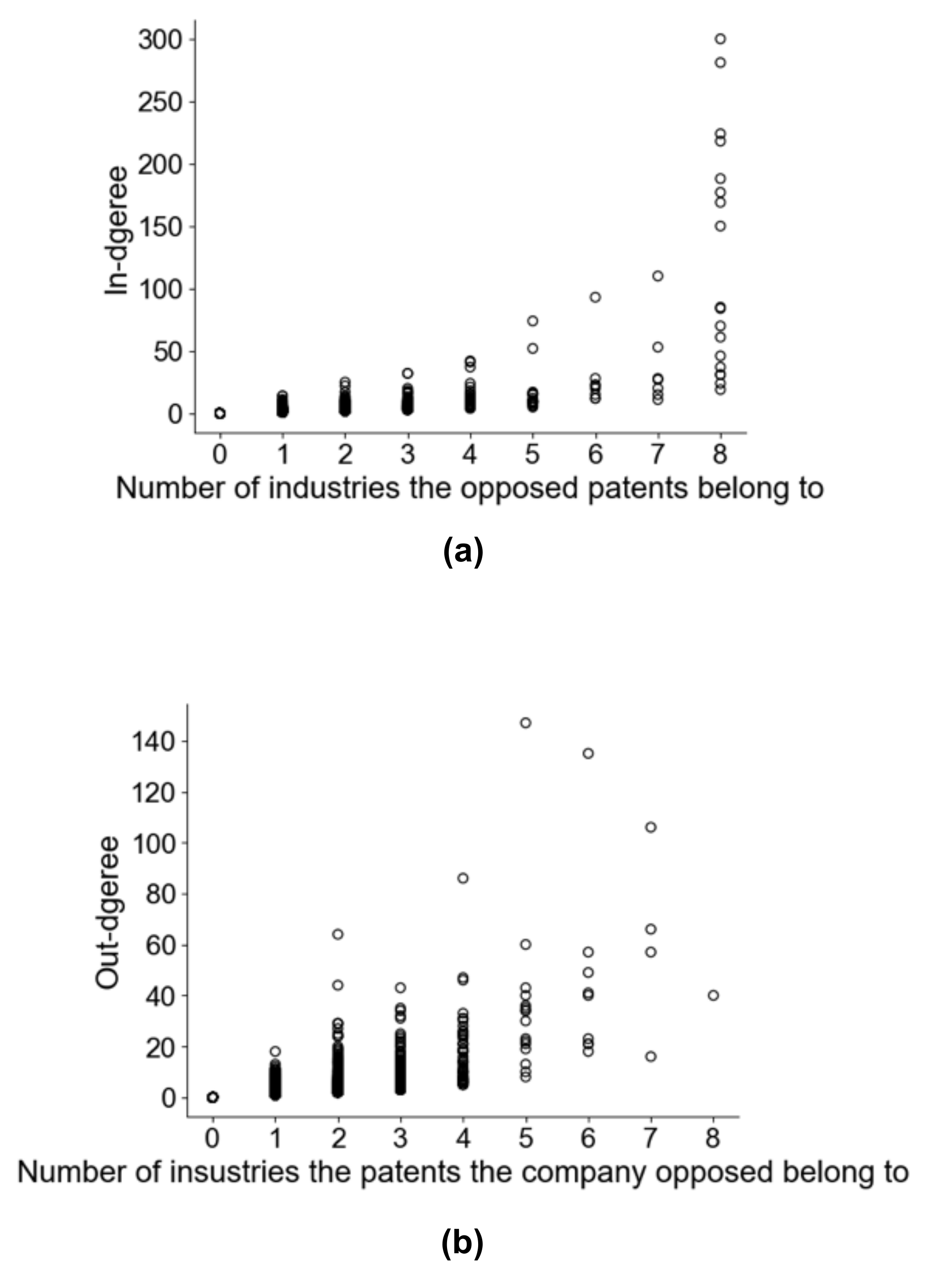}
      \caption{Patents' industrial sections and company's (a) in- and (b) out-degree distributions}
      \label{fig: industry-degree}   
      \end{center}
\end{figure}

Figure~\ref{fig: opposition network} visualises the opposition network.
There is one large connected component (displayed in the middle in the figure) and many small connected components (composed of up to 17 nodes, displayed in the periphery of the figure). A total of 7,489 companies are found in the largest component, which contains approximately 88\% of opposition edges. In the figure, each node's colour represents the geographical location of the company identified by the standard country code provided by the Orbis database. 
The existence of one large connected component suggests that, contrary to the expectations, opposition relationships are not bounded by the countries (or jurisdictions) in which companies are located, or industrial sections to which the patents belong. Regarding the country, we found only 26.6\% of the opposition edges that connect two companies located in the same country. The remaining 73.4\% of the edges connect companies located in different countries. Short of the between-country edges, the largest connected component breaks down into parts, and the largest connected component size becomes 1,617. Meanwhile, if we remove within-country edges and maintain between-country edges, 5,410 nodes are still connected. In other words, a majority of companies in the largest connected component have opposition relationships with those in different countries. 
Regarding the industrial section, Figure~\ref{fig: industry-degree}(a) shows the scatter plot of the in-degree of companies (i.e., the number of incoming edges the company has) and the number of industrial sections (up to 8, as described in Section~\ref{subsec:2.2}) to  which its owning patents belong. Likewise, Fig.~\ref{fig: industry-degree}(b) plots the out-degree of companies (i.e., the number of outgoing edges of the company) and the number of industrial sections to which the opposing patents belong. As shown in the figure, many companies own and oppose patents in various industrial sections. 
 
 Social networks with negative relationships are usually highly disconnected and do not have any clustering~\cite{Everett 2014}. The cross-country and cross-industry connections among companies make the opposition network distinct from such networks.

%
%

\subsection{Degree distribution and assortativity}
\label{subsec:3.2}
\begin{figure}[hbt]
\begin{center}
      \includegraphics[width=9cm]{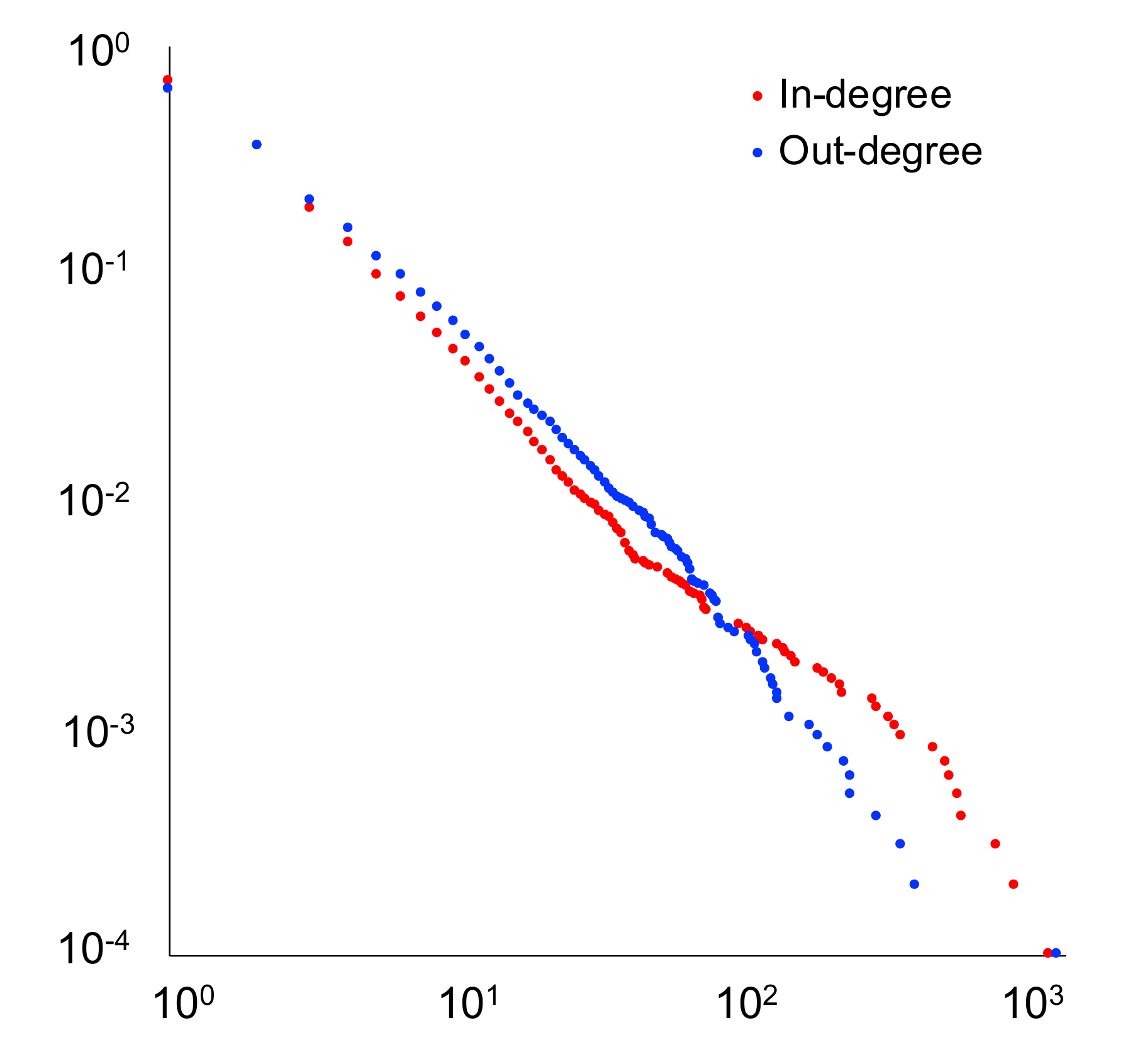}
      \caption{Complementary cumulative distributions of in-degrees and out-degrees of the opposition network}
      \label{fig: degree distribution}   
      \end{center}
\end{figure}

\begin{figure}[hbt]
\begin{center}
      \includegraphics[width=10cm]{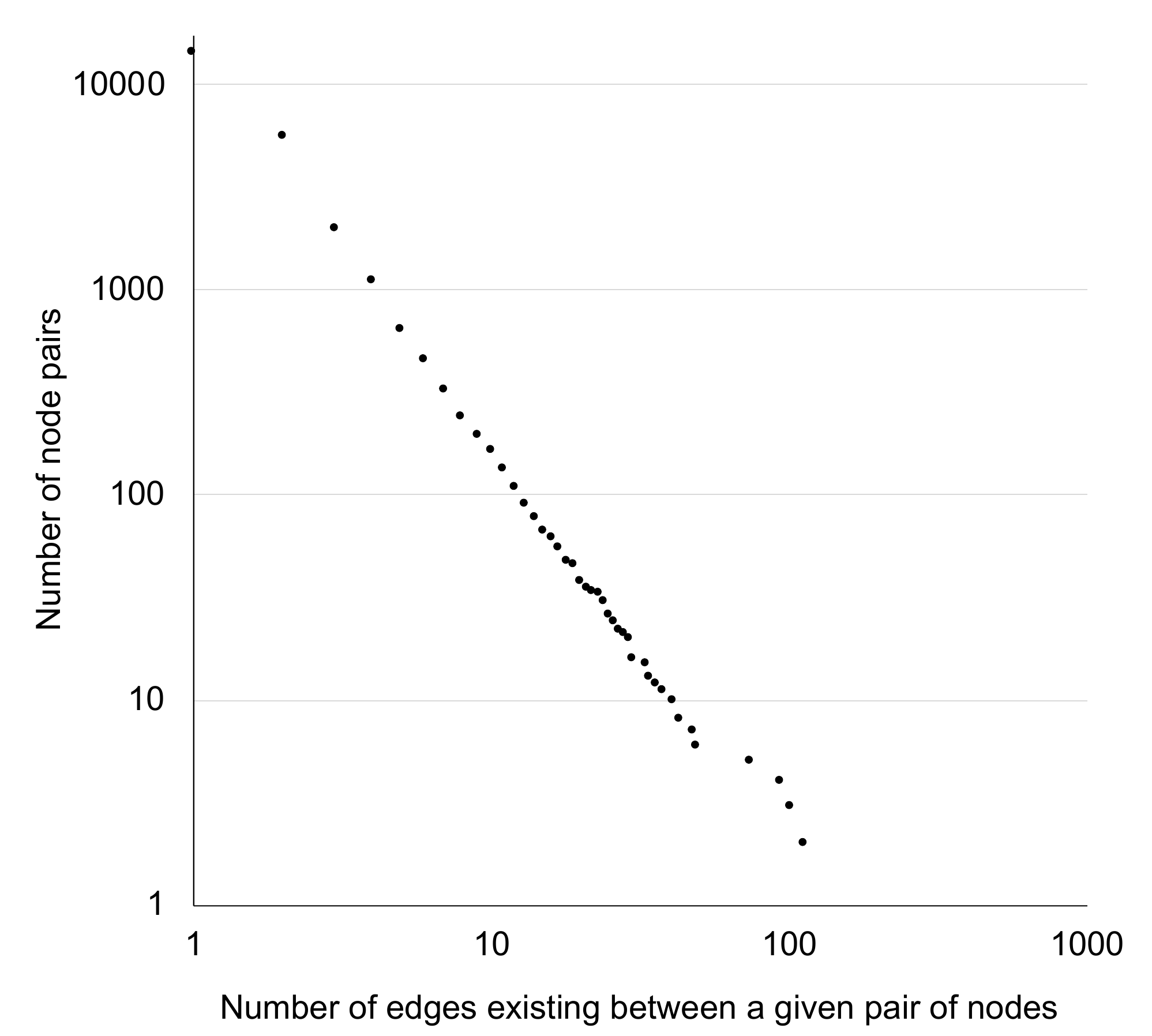}
      \caption{Destribution of the number of multiple edges}
      \label{fig: multiedge}   
      \end{center}
\end{figure}

Fig.~\ref{fig: degree distribution} shows the complementary cumulative distributions of the opposition network's in- and out-degrees, $P(x)$, evaluated from the raw data. The vertical axes show the proportion of companies that have a number of incoming/outgoing edges equal to or greater than the value given on the horizontal axis. Here, multiple edges were considered. In other words, the distributions are equivalent to the weighted degree distribution of the network in which multiple edges are disallowed and multiple occurrences of oppositions between a given pair of nodes are reflected as the weight of the edge. 
The distributions of both in- and out-degrees exhibit very heavy tails, meaning that a small number of nodes in the network hold the majority of the edges. Figure~\ref{fig: multiedge} shows the distribution of multiple edges. This distribution is also heavy-tailed, indicating that there are node pairs between which a considerable number of oppositions occurred. 

These results may imply that ``the rich get richer'' (the preferential attachment mechanism) works even for the formation of adversarial relationships in the inter-organisational network, which sets the opposition network aside from social negative relationship networks. Existing network formation models, including negative social relationships, to the best of our knowledge, are always based on mechanisms to balance triangles (see Ref.~\cite{Marvel 1771} and references within). 

We also calculated the degree correlation coefficient by ignoring the direction of edges~\cite{Newman 2002,Newman 2003}. 
The degree correlation coefficient is the Pearson correlation coefficient between the degrees found at the two ends of the same edge. This value $r$ varies between $-1\leq r \leq1$: the network is assortative if $r > 0$, and is disassortative if $r < 0$. Being assortative (or assortative mixing) means that large-degree nodes tend to form edges to each other and avoid connecting with small-degree nodes. The examination of various real-world networks suggested that social networks (such as co-authorship networks) exhibit assortative patterns, whereas technological and biological networks exhibit disassortative patterns~\cite{Newman 2002,Barabasi 2016}. Further, \cite{Newman 2003b} derives that assortativity is high in social networks based on the assumption that they can be described as projections of affiliation networks. This assumption is rather strong, which explains that many empirical social networks are actually disassortative.
Regarding the opposition network, we obtained the degree correlation coefficient $r = 0.16$, indicating its assortative pattern. In other words, the companies involved in many oppositions tend to have adversarial relationships with companies that similarly have many oppositions.
Evidently and mathematically, most finite-sized heavy-tailed distribution networks (including scale-free networks) are disassortative~\cite{Holme 2007}, with the exception of projected affiliation networks only (e.g., co-authorship networks)~\cite{Newman 2003b}. Therefore, we can assert that the opposition network exhibits a rather unusual type of network topology. 


\section{Motif analysis}
\label{sec:4}
Finally, we investigate the local relational patterns formed in the opposition network via the network motif analysis. 
Network motifs are small subgraphs that occur in a given network with significantly higher frequency than expected in the equivalent random networks (in terms of the numbers of nodes and edges, node degree distribution, etc.). Network motifs are small building blocks of a network, and are crucial to understanding the structure and functions of the network. Triads -- subgraphs of three nodes and the potential relationships among them, particularly, are considered to be structural foundations of social networks. Here, we mainly focus on the triadic relational motifs in the opposition network.

\subsection{Triadic motifs in the cumulative opposition network}
\label{subsec: triadic motifs}
First, we identified triadic motifs in the cumulative opposition network -- i.e., network in which directed edges represent opposition relationships that existed at some point during the period of study. The presence of multiple edges were neglected. In other words, if a company has opposed multiple patents owned by the same company, we simply concluded the presence of a directed opposition edge between them. The network was therefore directed and unweighted. We employed the motif counting algorithm method implemented in the graph-tool ~\cite{graph-tool}, which is a widely used Python module. 
The algorithm (like many other motif analysis algorithms) measures the statistical significance of the occurrence of each of the possible triadic relational patterns in a given network. The significance measure, Z-score, is defined as the difference of the frequency of the motif in the target network and its mean frequency in a set of randomised networks, divided by the standard deviation of the frequency values for the randomised networks~\cite{Milo 2002}. We used 100 randomised networks as null-models generated by reshuffling edges while retaining the degree sequence of the network. 

\begin{figure}
\begin{center}
      \includegraphics[width=11.8cm]{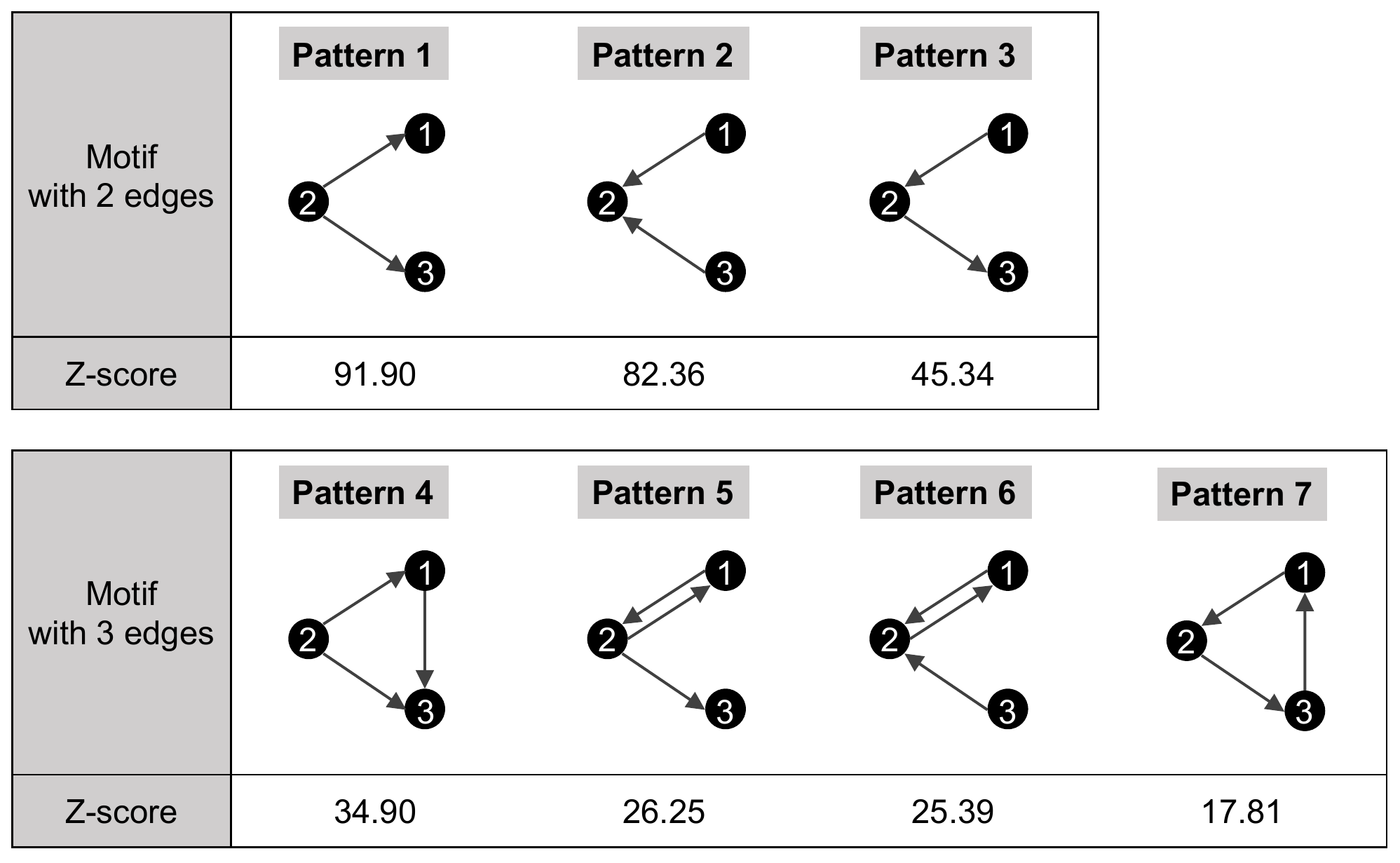}
      \caption{Triadic motifs identified in the opposition network}
      \label{fig: triadic motifs}   
      \end{center}
\end{figure}

Figure~\ref{fig: triadic motifs} shows the results. In the figure, the identified triadic motifs (Patterns 1 to 7) are illustrated with their Z-scores. Pattern 1 (i.e., a significantly large number of companies oppose more than one company) and Pattern 2 (i.e., a significantly large number of companies have patents opposed by more than one company) are apparent from the in- and out-degree distributions (see Fig.~\ref{fig: degree distribution}). 
Pattern 3 suggests that a chain of opposition relationships also occurs frequently. Companies involved in an opposition chain then tend to oppose each other too (Patterns 5 and 6).

The frequent appearance of negative relationship triangles such as Pattern 4 indicates the high tendency of the occurrence of ``the enemy of my enemy is also my enemy'' situation. This contradicts the low level of transitivity in social negative relationship networks (i.e., ``enemies of enemies are not necessarily enemies'')~\cite{Everett 2014}. We further investigate how such negative triads may emerge over time.


\subsection{Temporal motifs}
\label{subsec: temporal motifs}
\begin{figure}
\begin{center}
      \includegraphics[width=9cm]{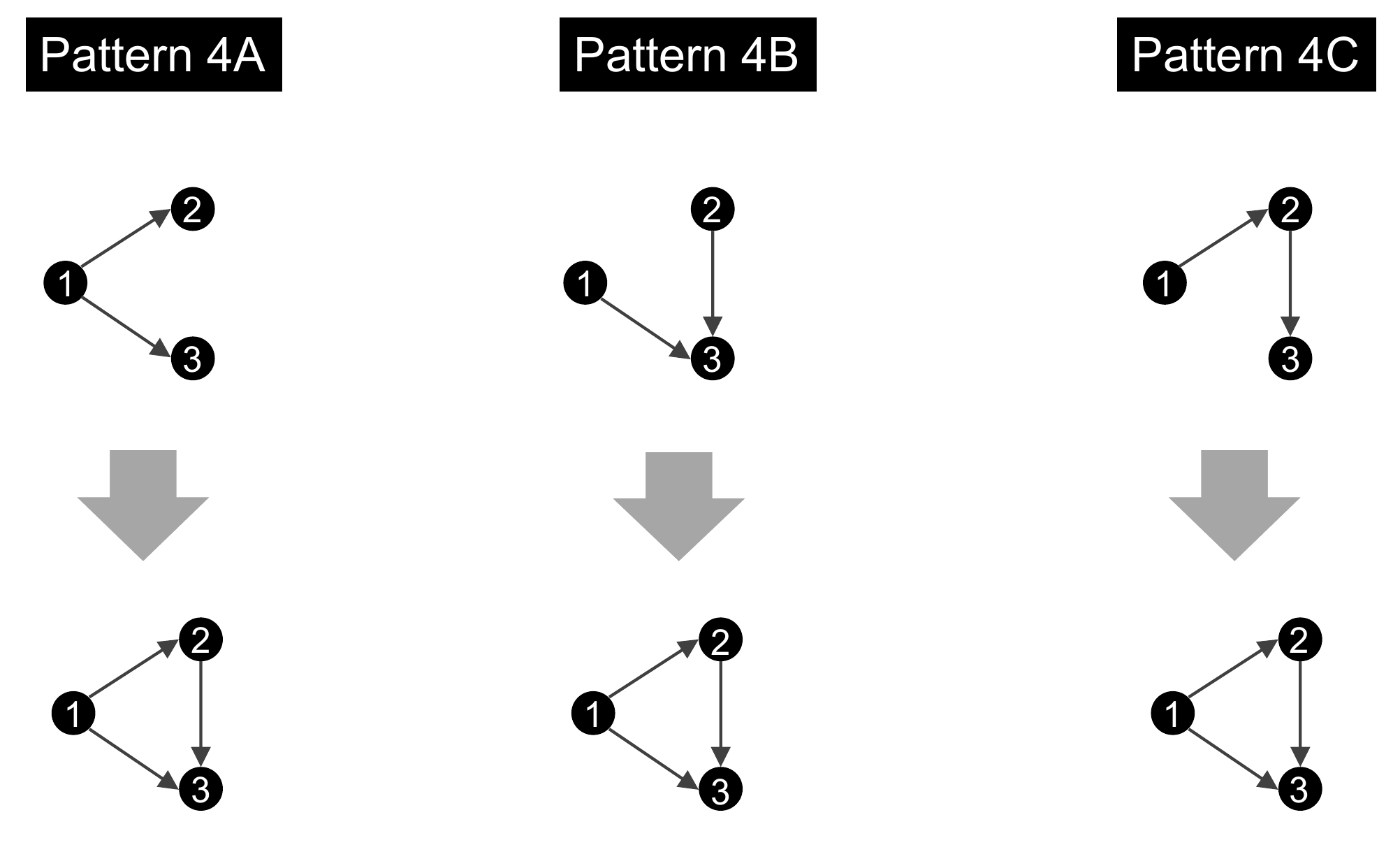}
      \caption{Temporal patterns of the occurrence of Pattern 4 in Fig.~\ref{fig: triadic motifs}
        \protect \newline {\it Note: The three open triads shown above the arrows can turn to Pattern 4 (shown below the arrows) by forming one more edge over time.}}
         \label{fig: temporal pattern possibility}  
         \end{center}
\end{figure}

Pattern 4 can occur by adding another opposition edge to Patterns 1, 2 and 3. In other words, one can make the following three possible logical interpretations for the emergence of Pattern 4 (see Fig.~\ref{fig: temporal pattern possibility}):
\begin{itemize}
    \item Two companies that had been opposed by the same company formed an adversarial relationship (Pattern 4A).
    \item Two companies that had opposed the same company formed an adversarial relationship (Pattern 4B).
    \item  A company got opposed by its opposer's opposer (two steps away) (Pattern 4C).
\end{itemize}

\begin{figure}
\begin{center}
      \includegraphics[width=14cm]{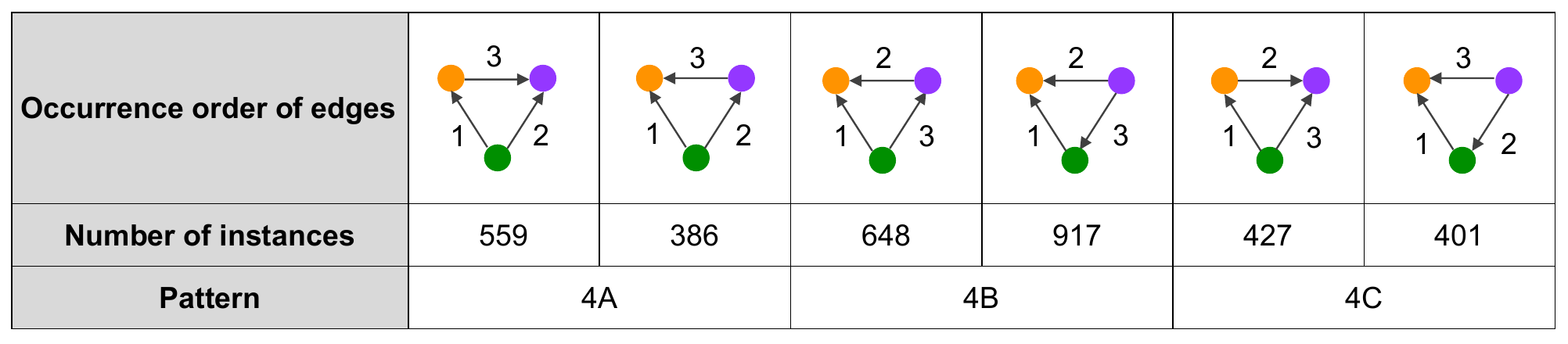}
      \caption{Number of instances of Patterns 4A, 4B and 4C in the opposition network 
       \protect \newline {\it Notes: The edge labels of the temporal triadic motifs correspond to the ordering of the edges. Patterns 4A to 4C  (shown in Fig.~\ref{fig: temporal pattern possibility}) do not consider the ordering of the first 2 edges. For example, both the extreme left and second extreme left motifs impel the formation of Pattern 4A.}}
      \label{fig: temporal triadic motifs}   
      \end{center}
\end{figure}

Such temporal patterns of the occurrence of edges can be identified by applying the temporal motif counting method. Various such methods have been proposed (e.g., \cite{Kovanen 2011}), with different algorithms and definitions of temporal motifs. In our analysis, the matter of concern is the triadic closure process. In other words, we examine which patterns among Patterns 4A, 4B, and 4C may be a more plausible logic than others. In the process of the formation of Pattern 4A, nodes may be involved in other oppositions. For example, node 1 in Pattern 4A, shown in Fig.~\ref{fig: temporal pattern possibility}, may have opposed another node outside this triangle before the formation of the final edge between nodes 2 and 3. Otherwise, node 1 may have opposed node 2 repeatedly. We considered all of such cases contributing to the emergence of Pattern 4A. In other words, our aim was to ensure that every occasion that edges form a particular pattern within the observation period is counted. Therefore, we employed the method provided by the Stanford Network Analysis Platform (SNAP)~\cite{Paranjape 2017,SNAP}, which serves this purpose. 
Figure~\ref{fig: temporal triadic motifs} shows the number of instances (i.e., the actual counts) of the occurrence of each temporal triadic pattern, counted using this method. Each of Patterns 4A, 4B, and 4C can occur in two ways if the ordering of the first 2 edges is taken into account. Thus, the number of instances of Pattern 4A, for example, is the sum of 559 and 386 (= 945). 

We found that, compared to Pattern 4A (945 times) and Pattern 4C (827 times), Pattern 4B occurred considerably more frequently. In other words, two companies that have opposed the same company tend to form an adversarial relationship. This may imply that companies that have a common enemy share technological interests and thus are prone to becoming rivals. Indeed, the assertion that enemy of my enemy may also be my enemy holds trues.


\subsection{Motifs of opposition and collaboration relationships}
\label{subsec: collaboration}
In the opposition network, we concluded that the enemies of my enemy tend to be my enemies. Nevertheless, companies also have collaborative relationships. In this section, we investigate where in the opposition network collaboration relationships occur and how the collaboration relationships would interact with opposition relationships. 

First, we focus on dyads. We consider a relationship between two companies adversarial when one company opposes the other. They may form a reverse opposition edge (as shown in Fig.~\ref{fig: triadic motifs} for the case of triads). Otherwise, their relationship remains adversarial (or may become more neutral over time). In general, the relationship between a pair of companies does not alternate between collaboration and rivalry. 
However, we found some cases in which an adversarial relationship turned to a collaborative relationship. We counted the number of instances in which an opposition edge appeared at time $t_1$ and a collaboration edge appeared between the same pair of nodes at $t_2$ ($t_1 < t_2$). 
There are 14,320 unique pairs of nodes connected via one or more opposition edges. Among those, 374 pairs formed a reverse opposition edge, while 9 pairs formed a collaborative relationship later. Although very uncommon, it is possible for companies to transition from an adversarial to a collaborative relationship. 
Meanwhile, we found 14 out of the 1,554 unique pairs of nodes that were in a collaborative relationship having formed an opposition relationship later in time. 
Companies very rarely change their relationships with others, from rivalry to collaborative or vice versa. Further, if nothing else, there is a higher tendency for collaboration to turn to rivalry. 

\begin{figure}
\begin{center}
      \includegraphics[width=14cm]{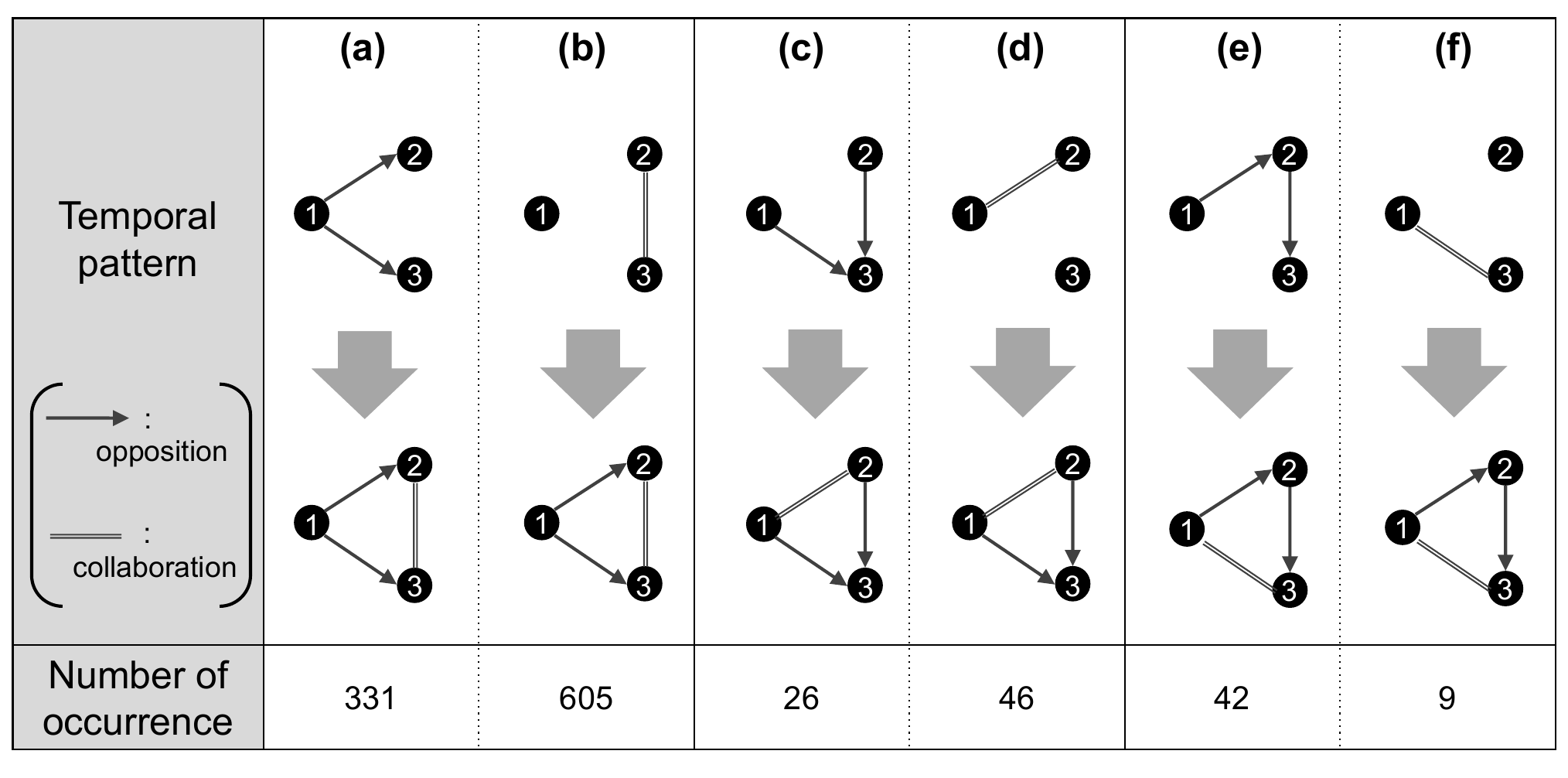}
      \caption{Temporal patterns of the occurrence of triads with two opposition relationships and one collaboration relationship
        \protect \newline {\it Notes: Each panel corresponds to a temporal triadic pattern formed by two opposition edges and one collaboration edge that occurs in a specific order. The occurrence order of the two opposition edges was disregarded.}}
      \label{fig: collaboration triads}   
      \end{center}
\end{figure}

Second, We moved onto the triads. As already discussed, Pattern 4 is formed by the triadic closure of Patterns 1--3 (i.e., by adding an opposition edge between the unconnected pair of nodes), all of which evidently occur more frequently than expected (see Fig.~\ref{fig: triadic motifs}). Pattern 4 is structurally imbalanced. The triangle becomes structurally balanced if the third edge added to Patterns 1--3 is a collaboration rather than an opposition edge. 
We further counted the temporal occurrence of such balanced triads while focussing on whether a collaboration edge was formed before or after two opposition edges. Figure~\ref{fig: collaboration triads} summarises the results. Pattern 1 with a collaboration edge (hereafter, Pattern 1') can occur in the following two ways: two opposition edges occur before a collaboration edge, or vice versa, as shown in (a) and (b). Likewise, (c) and (d) are for Pattern 2 with a collaboration edge (hereafter, Pattern 2'), and (e) and (f) are for Pattern 3 with a collaboration edge (hereafter, Pattern 3'), respectively. The number of each pattern's occurrences is shown in the bottom row of the figure. For example, when a company had opposed the two companies, the case where these two companies started collaborating at any point in time after that occurred 331 times (see the panel (a)). 

As the figure suggests, Pattern 1' occurred considerably more frequently (331 + 605 = 936 times) than Patterns 2' and Pattern 3'. In other words, the situation in which two rivals --or ``enemies''-- of a company collaborate together (= Pattern 1') is more likely to occur compared to the situation in which collaborating companies share a common rival (= Pattern 2') or a rival of a rival of a company is its collaborator (= Pattern 3'). 
When we looked further into the emergence of Pattern 1', the pattern (a) occurred considerably less frequently than the pattern (b). This suggests that being opposed by the same company (i.e., node 1 in (a))  does not necessarily make two companies (i.e., nodes 2 and 3) sharing technological knowledge or a common interest, which is reasonable given that many companies engage in patenting in multiple industrial sections. 

In sum, we did not observe any sign of interactions between collaboration and opposition relationships. The mechanism for the emergence of opposition relationships is clearly different from that of collaboration relationships, underlining the necessity of further studies on the opposition network.

\section{Conclusion}
\label{sec:5}
We constructed the opposition network in which nodes represent companies and directed edges represent opposition relationships. 
This study is the first to map the topology of the inter-organisational patent opposition network. This study is novel not only as a study of patents but also as a study of networks with negative relationships. 

The results of our analysis elucidated the characteristics of this network that are distinct from other negative relationship networks discussed in the social network analysis. 
A large number of companies are interconnected, suggesting that the formation of opposition relationships is not bounded by the country or industrial section. The coexistence of a heavy-tailed degree distribution and the assortative mixing by degree makes the opposition network topology a rare type, highlighting the need to be careful when applying knowledge derived from analyses of social negative relationship networks. 
This need was further enhanced by the identification of structurally imbalanced triads overrepresented in the network. The disagreement between the significant local structural patterns in the opposition network and those in social networks suggests that the mechanisms of the emergence of this network would differ from social negative relationship networks.
Fundamentally, inter-organisational relationships are formed strategically. One feature that characterises opposition relationships is that a high probability of opposition indicates a high value of a patent. An inter-organisational adversarial relationship captured by patent opposition may, therefore, be considerably different from social negative relationships, such as inter-personal dislike.

The future direction of this research is to further investigate how opposition networks are formed and relate the identified network characteristics to the knowledge on patenting strategies. A more detailed analysis of patent attributes (e.g., industrial sections, registered patent office, patent family, closely related inventions, etc.) is also necessary in order to enrich the understanding of the complex network of the patenting strategies of companies.

\section*{Funding}
This work was supported in part by the JSPS KAKENHI (19K04893).


 
 
 


\end{document}